\documentclass{aa}
\usepackage{graphicx}
\usepackage{txfonts}
\begin{document}
\title{Radio and submillimetre observations of wind structure in $\zeta$~Pup}


\author{R. Blomme \inst{1} \and G.C. Van de Steene \inst{1} \and 
        R.K. Prinja \inst{2} \and M.C. Runacres \inst{1} \and
        J.S. Clark \inst{2}
          }

\offprints{R. Blomme, \email{Ronny.Blomme@oma.be}}

\institute{Royal Observatory of Belgium,
           Ringlaan 3, B-1180 Brussel, Belgium 
         \and
             Department of Physics \& Astronomy, University College London, 
             Gower Street, London WC1E 6BT, UK
             }

   \date{Received date; accepted date}

   \abstract{
We present radio and submillimetre observations of the
O4I(n)f star $\zeta$~Pup, and discuss 
structure in the outer region of its wind ($\sim 10-100~R_*$).
The properties of bremsstrahlung, the dominant emission process at these 
wavelengths, make it sensitive to structure and allow us to 
study how the amount of structure changes in the wind by comparing the fluxes
at different wavelengths.
Possible forms of structure at these distances include
Corotating Interaction Regions (CIRs), stochastic clumping, a disk or a polar
enhancement. As the CIRs are azimuthally asymmetric, they should result
in variability at submillimetre or radio wavelengths.
To look for this variability, we acquired 3.6 and 6 cm
observations with the Australia Telescope Compact
Array (ATCA), covering about two rotational
periods of the star. We supplemented these with archive observations
from the NRAO Very Large Array (VLA), which cover a much longer time scale.
We did not find variability at more than the $\pm 20$~\% level.
The long integration time does allow an accurate determination
of the fluxes at 3.6 and 6 cm.
Converting these fluxes into a mass loss rate, we find 
$\dot{M}$ = $3.5 \times 10^{-6}$ $M_{\sun}/\mathrm{yr}$.
This value confirms the significant discrepancy with the mass loss
rate derived from the H$\alpha$ profile, making $\zeta$~Pup an
exception to the usually
good agreement between the H$\alpha$ and radio mass loss rates.
To study the run of structure as a function of distance, we supplemented
the ATCA data by observing $\zeta$~Pup at 850 $\mu$m with the 
James Clerk Maxwell Telescope (JCMT) and at 20 cm with the VLA.
A smooth wind model shows that the millimetre fluxes are too high compared to the
radio fluxes. While recombination of helium in the outer wind cannot be discounted
as an explanation, the wealth of evidence for structure strongly suggests
this as the explanation for the discrepancy. Model calculations show
that the structure needs to be present in the inner $\sim 70~R_*$
of the wind, but that it decays significantly, or maybe even
disappears, beyond that radius.
   }

   \maketitle
\section{Introduction}

\object{Zeta Puppis} (\object{HD 66811}; O4I(n)f) 
is one of the most-studied O-type stars.
Its stellar wind is driven by radiation pressure, whereby the momentum
of the stellar photons is transferred to the wind material
via line opacity.
With other early-type
stars it shares the property that its wind is structured. Evidence
for this structure for O-type stars in general
is seen in the Discrete Absorption Components (DACs)
and black troughs in the ultraviolet resonance 
lines (e.g. Prinja et al.~\cite{Prinja+al90}), the presence of
X-ray emission (Sciortino et al.~\cite{Sciortino+al90})
and the excess flux at infrared and millimetre 
wavelengths (Runacres \& Blomme~\cite{Runacres+Blomme96}).
Specifically for \object{$\zeta$~Pup}, 
the significant discrepancy between the mass loss 
rate derived from the H$\alpha$ spectral line and that from the radio continuum 
flux (Petrenz \& Puls~\cite{Petrenz+Puls96})
also points to structure in the wind.

Various types of structure have been proposed to explain the observations.
Inhomogeneities on the stellar surface (e.g., due to non-radial pulsations),
or magnetic fields,
result in somewhat different radiative forces, thereby
creating fast and slow streams of gas. As these streams also rotate,
they collide, creating
Corotating Interaction Regions (CIRs). These are large-scale,
spiral-shaped structures
in the wind that corotate with the stellar surface. In some cases,
the CIRs can explain the
most interesting property of DACs: the recurrence time scale is in agreement
with the estimated rotation period
(Mullan~\cite{Mullan86}, Cranmer \& Owocki~\cite{Cranmer+Owocki96}). 

Another type of structure that can exist in the wind is of a smaller-scale,
stochastic type. It is caused by the inherent instability of the radiative
driving mechanism (Owocki~\cite{Owocki00}). 
Stochastic structure has been used to explain the
X-ray fluxes (Lucy~\cite{Lucy82a}, Hillier et al.~\cite{Hillier+al93},
Feldmeier et al.~\cite{Feldmeier+al97}), 
the black troughs in the UV resonance lines
(Lucy~\cite{Lucy82b})
and the excess millimetre flux (Blomme et al.~\cite{Blomme+al02}).

Finally, it should be noted that $\zeta$~Pup is a rapid rotator
(at 43~\% of critical velocity).
This suggests the possibility
that $\zeta$~Pup has a disk,
or at least some enhancement of its density near the equatorial plane.
Harries \& Howarth~(\cite{Harries+Howarth96}) derived a density contrast
of at least 1.3 from their linear spectropolarimetry of H$\alpha$.
Such equatorial density enhancement has been used 
(Petrenz \& Puls~\cite{Petrenz+Puls96}) to explain the
discrepancy that smooth and spherically symmetric models give for
the mass loss rates derived from H$\alpha$ and
radio observations.

While observational evidence for the existence of disks in certain stars
(e.g. Be stars) is strong, there are theoretical difficulties in
understanding how these disks form (Owocki et al.~\cite{Owocki+al98}).
One of the explanations proposed was the
wind-compressed disk model (Bjorkman \& Cassinelli~\cite{Bjorkman+Cassinelli93}),
but a more detailed study showed that
the non-radial component of the line force and gravity darkening
can combine to inhibit the formation of the disk. Interestingly, 
in some circumstances, the detailed models show
an enhancement of the density towards the poles rather than the equator
(Owocki et al.~\cite{Owocki+al96}, \cite{Owocki+al98},
Petrenz \& Puls~\cite{Petrenz+Puls00}).

Observations at submillimetre and radio wavelengths are relevant to
structure. This is
because bremsstrahlung, the dominant emission process at these wavelengths,
has two interesting properties.
First, its opacity is proportional to the {\em wavelength}
squared.
Therefore, the observed emission originates above a characteristic
radius that increases with wavelength (typical values are $\sim 5~R_*$
at 1 mm and $\sim 100~R_*$ at 20 cm). Secondly, free-free emission
also depends on the {\em density} squared. This makes it a good indicator
of structure: a structured wind will have more radio emission
than a smooth one.
By looking for variability at a certain wavelength, one can hope to distinguish
between structure that has azimuthal symmetry (stochastic, disk or polar outflow)
and structure that lacks azimuthal symmetry (CIRs).
By looking at increasing wavelengths, one can see how the ``amount" of structure
changes in the wind.

We have already applied these techniques
to the submillimetre and radio observations
of the B0Ia star \object{$\epsilon$~Ori} (Blomme et al.~\cite{Blomme+al02}).
For this star, we did not detect variations in the radio emission
(at the 25~\% level),
but we found that
the millimetre flux is substantially higher than a smooth wind
model predicts. This discrepancy was interpreted with
a model for stochastic structure, showing that 
considerable structure must persist up to at least $\sim 10~R_*$ in the
wind of $\epsilon$~Ori. The present paper extends the study of structure
in the outer wind to a luminous early O-type star.

To look for variability at radio wavelengths, we acquired 3.6 and 6 cm
observations with the Australia Telescope Compact 
Array\footnote{The Australia Telescope is funded by the Commonwealth of
              Australia for operation as a National Facility managed by
              CSIRO.}
(ATCA), during a 12-day observing session, which covers about two rotational
periods of the star. To see how structure changes as a function of
distance, observations were collected at 850 $\mu$m with the James Clerk Maxwell 
Telescope\footnote{The JCMT is operated by the Joint Astronomy Centre in Hilo,
              Hawaii on behalf of the parent organizations Particle Physics and
              Astronomy Research Council in the United Kingdom, the National
              Research Council of Canada and The Netherlands Organization for
              Scientific Research.}
(JCMT), and at 20 cm with the NRAO Very Large 
Array\footnote{The National Radio Astronomy
             Observatory is a facility of the National Science Foundation
             operated under cooperative agreement by Associated Universities,
             Inc.}
(VLA).
We supplemented this material with data from the VLA archive.
A log of all data used in this paper is given in
Table~\ref{table data zeta pup}. 
Upper limits from other observations (not discussed here)
are listed in Wendker~(\cite{Wendker95}).

\begin{table}[t]
\caption{New observations and archive data of $\zeta$~Pup.}
\label{table data zeta pup}
{\small
\begin{tabular}[]{lllllllllllll}
\\
\multicolumn{2}{c}{programme} & \multicolumn{1}{l}{date}     & \multicolumn{1}{l
}{$\lambda$} \\
\multicolumn{2}{c}{name} & & \multicolumn{1}{l}{(cm)} & \\
\hline
\multicolumn{3}{l}{{\em New observations}}\\
\hline
\multicolumn{2}{l}{ATCA}\\
& C824& 1999-09-15$\rightarrow$27 & 3.6+6      & \\
\multicolumn{2}{l}{VLA}\\
& AB1017& 2002-03-27 & 20         & \\
\multicolumn{2}{l}{JCMT}\\
& M00BU20& 2000-10-12 & 0.0850  &\\
\hline
\multicolumn{3}{l}{{\em Archive data}}\\
\hline
\multicolumn{2}{l}{VLA}\\
& FLOR  & 1978-07-23 & 6          & \\
& FLOR  & 1978-10-13 & 6          & \\
& BIEG  & 1978-11-05 & 6          & \\
& NEWE  & 1979-02-09 & 2          & \\
& FLOR  & 1979-02-16 & 20         & \\
& CHUR  & 1979-07-12 & 6          & \\
& FLOR  & 1981-10-18 & 1.3+2+6+20 & \\
& AA28  & 1984-03-07 & 2+6        & \\
& AB327 & 1985-01-29 & 2+6        & \\
& AH365 & 1989-05-13 & 3.6        & \\
& AC308 & 1995-01-17 & 20         & \\
\hline
\end{tabular}
}
\end{table}

In Sect.~\ref{section stellar parameters}, we present the
stellar parameters of $\zeta$~Pup.
The ATCA, VLA and JCMT observations are discussed in 
Sects.~\ref{section ATCA observations}, \ref{section VLA observations} and
\ref{section JCMT observations} respectively.
The interpretation of the observational material
is discussed in Sect.~\ref{section discussion} and conclusions are drawn in
Sect.~\ref{section conclusions}.

\section{The parameters of $\zeta$~Pup}
\label{section stellar parameters}

Table~\ref{table stellar parameters} lists the $\zeta$~Pup parameters
we use in this paper.
The stellar parameters and wind parameters are taken from 
the unified NLTE model of Puls et al.~(\cite{Puls+al96}).
These values are in close agreement with models
by Bohannan et al.~(\cite{Bohannan+al90}) and
Pauldrach et al.~(\cite{Pauldrach+al94}).
The traditionally assumed distance of 450 pc 
(Kudritzki et al.~\cite{Kudritzki+al83}) falls well within the
Hipparcos error bar, so we will use $d=450$ pc throughout this paper.
The radius of 19 $R_{\sun}$ at a distance of 450 pc corresponds
to an angular diameter of 0.39 milli-arcsec (mas), in good agreement with 
interferometric measurements ($\theta_{\rm LD} = 0.42 \pm 0.03$ mas;
Hanbury Brown et al.~\cite{HanburyBrown+al74}).

\begin{table}
\caption{Stellar parameters of $\zeta$~Pup.}
\label{table stellar parameters}
\begin{tabular}{llll}
parameter & value & reference \\
\hline\\
RA (J2000)             & $08^{\rm h}03^{\rm m}35{\fs}0467$ & SIMBAD \\
Dec (J2000)            & $-40{\degr}00{\arcmin}11{\farcs}332$ & \quad catalogue \\
$\mu_\alpha {\rm cos}~\delta$ & $-31.7 \pm 0.5$ mas/yr & G01 \\
$\mu_\delta$           & $+17.6 \pm 0.6$ mas/yr & G01 \\
$V$ magnitude          & 2.25 & M87 \\
$B-V$                  & $-0.27$ & M87 \\
$E_{\rm B-V}$          & 0.044 & S77 \\
spectral type          & O4I(n)f & W72 \\
$T_{\rm eff}$          & 42000 K & P96 \\
log g                  & 3.60 & P96 \\
$\log L/L_{\sun}$      & 6.00 & P96 \\
$R_*$                  & $19 R_{\sun}$ & P96 \\
$M_*$                  & $52.5 M_{\sun}$ & P96 \\
$N_{\rm He}/N_{\rm H}$ & 0.12 & P96 \\
$v \sin i$             & 220 km/s & P96 \\
$v_\infty$             & 2250 km/s & P96 \\
$\beta$                & 1.15 & P96 \\
$\dot{M}$              & $5.9 \times 10^{-6} M_{\sun}/\mathrm{yr}$ & P96, 
                                                      from H$\alpha$ \\
$d$                    & 450 pc & K83 \\
                       & $429^{+120}_{-77}$ pc & Hipparcos \\
\\
\hline
\end{tabular}\\
References:\\
\begin{tabular}{ll}
G01  & Gontcharov et al.~(\cite{Gontcharov+al01}) \\
K83  & Kudritzki et al.~(\cite{Kudritzki+al83}) \\
M87  & average from Mermilliod~(\cite{Mermilliod87}) \\
P96  & Puls et al.~(\cite{Puls+al96}) \\
S77  & Snow et al.~(\cite{Snow+al77}) \\
W72  & Walborn~(\cite{Walborn72}) \\
\end{tabular}
\end{table}

$\zeta$~Pup shows considerable evidence for structure in its stellar wind.
Discrete Absorption Components (DACs) are seen to move through the
ultraviolet resonance lines (Prinja et al.~\cite{Prinja+al92} and
references therein). 
In 1995, $\zeta$~Pup was one of three targets
observed continuously during 16 days by the {\em International Ultraviolet
Explorer} (IUE Mega Campaign, Massa et al.~\cite{Massa+al95}).
The analysis of this high-quality dataset by 
Howarth et al.~(\cite{Howarth+al95}) reveals a 19.2 h and a 5.2 day period.
The 19.2 h period is the recurrence time of the DACs.
The 5.2 day period indicates modulation on a global scale, and is in good 
agreement with the rotation period estimated from $v \sin i$, suggesting 
that material in the wind corotates with the star.
One possible cause for this corotating material
is the presence of a (weak) magnetic field. Attempts to detect the
magnetic field for $\zeta$~Pup have so far yielded a null result with an
uncertainty of 100-200 G 
(Barker et al.~\cite{Barker+al81},
Chesneau \& Moffat~\cite{Chesneau+Moffat02}).

$\zeta$~Pup also shows variability in the H$\alpha$ and He {\sc ii} 4686
line profiles
(Moffat \& Michaud~\cite{Moffat+Michaud81} and references therein,
Hendry \& Bahng~\cite{Hendry+Bahng81},
Reid \& Howarth~\cite{Reid+Howarth96},
Eversberg et al.~\cite{Eversberg+al98}).
The presence of variability in these wind-formed lines is consistent
with large-scale structures corotating through the wind.
Eversberg et al. however interpret their He {\sc ii} 4686 observations
in terms of smaller-scale structures (``clumps") moving out in the wind.
Fullerton et al.~(\cite{Fullerton+al96}) and
Reid \& Howarth~(\cite{Reid+Howarth96}) detected variability in a number
of optical lines (He {\sc i}, He {\sc ii}, N {\sc iv}, C {\sc iv}). 
The variations have a period of 8.54~h. 
Both papers attempt to interpret the
variations in terms of non-radial pulsations, but they also raise concerns
that these lines might be wind contaminated.
The photospheric perturbations due to non-radial pulsations are another
possible way of creating CIRs.

Optical continuum fluxes also show variability:
the 5.2 day rotation period was detected in optical photometry
(Balona~\cite{Balona92}). From Hipparcos photometry,
Marchenko et al.~(\cite{Marchenko+al98}) derive a period which is half
the rotation period.

Other indicators of structure are the continuum flux excess at
infrared and millimetre wavelengths 
(Runacres \& Blomme~\cite{Runacres+Blomme96}) and 
the presence of X-ray emission
(Long \& White~\cite{Long+White80}). Models by 
Hillier et al.~(\cite{Hillier+al93})
and Feldmeier et al.~(\cite{Feldmeier+al97}) 
show that structure can explain the
X-ray emission. Earlier claims that the observed X-ray emission is variable
(Collura et al.~\cite{Collura+al89}) were later shown to be incorrect
(Bergh\"ofer \& Schmitt~\cite{Berghoefer+Schmitt94}). 
Later, Bergh\"ofer et al.~(\cite{Berghoefer+al96})
found variability in the ROSAT X-ray data (on roughly the same
time scale as the H$\alpha$ variability), but this
was not confirmed by the ASCA data (Oskinova et al.~\cite{Oskinova+al01}).

\section{ATCA observations}
\label{section ATCA observations}

\subsection{Data}

We used the Australia Telescope Compact Array (ATCA) for a long observing
session on $\zeta$~Pup in September 1999. The $\sim$~27~h on-target integration
time was divided into 8 runs, spread over 12 days, thus providing
coverage of the $\sim$~5.2 day rotation period. 
A detailed observing log is presented in Table~\ref{table ATCA}.
At the time of the observation, ATCA was in configuration 6A, with the
longest baseline at 5939~m and the shortest one at 337~m.
The continuum observations were done simultaneously at 3.6~cm
(X-band; 8.688~GHz) and 6~cm (C-band; 4.848~GHz). 
The observing bandwidth was 128 MHz.
A single observing run consists
of repetitively observing $\zeta$~Pup for 15~min and then the phase
calibrator \object{PKS B0823-500} for 5~min.  The flux calibrator 
\object{PKS B1934-638}
 was observed when possible at the beginning or end of the run.
The flux calibrator could not be observed during the fourth (SEP20) and seventh
observing run (SEP25) due to scheduling or technical problems.

\begin{table*}
\caption[]{ATCA data for $\zeta$~Pup. Start and end times of each observing run
are given, as well as the total duration of the run and the integration
time on $\zeta$~Pup. The measured fluxes at 3.6 and 6~cm are listed, with
their error bars and the size of the beams (major $\times$
minor axis and position angle).
The data obtained on SEP23-24 were split in two sets in order to have the
same time span as in the other observing runs.
}
\label{table ATCA}
\begin{tabular}{lllllllr@{.}llll}
\hline
\multicolumn{1}{c}{start date-time} & \multicolumn{1}{c}{end date-time} & 
   \multicolumn{1}{l}{total} & \multicolumn{1}{l}{$\zeta$~Pup} &
   \multicolumn{1}{c}{F$_{\rm 3.6~cm}$} & \multicolumn{1}{c}{F$_{\rm 6~cm}$} & 
   \multicolumn{1}{c}{beam$_{\rm 3.6~cm}$} & \multicolumn{2}{c}{PA}  & 
   \multicolumn{1}{c}{beam$_{\rm 6~cm}$} \\
\multicolumn{1}{c}{1999-} & \multicolumn{1}{c}{1999-} & 
   \multicolumn{1}{c}{time (h)} & \multicolumn{1}{c}{time (h)} & 
   \multicolumn{1}{c}{(mJy)} & \multicolumn{1}{c}{(mJy)} & 
   \multicolumn{1}{c}{(\arcsec $\times$ \arcsec)} & \multicolumn{2}{c}{(deg)} & 
   \multicolumn{1}{c}{(\arcsec $\times$ \arcsec)} \\
\hline 
SEP15-23:47:55 & SEP16-03:32:15 & 3.30 & 2.44 &
 1.98 $\pm$ 0.12 & 1.40 $\pm$ 0.11 & 4.05$\times$0.68 &  33&6 & 7.43$\times$1.23 \\
SEP18-23:07:15 & SEP19-03:01:15 & 3.52 & 2.56 &
 2.10 $\pm$ 0.12 & 1.49 $\pm$ 0.11 & 3.78$\times$0.67 &  29&4 & 6.96$\times$1.20 \\
SEP19-22:09:05 & SEP20-01:59:35 & 3.44 & 2.50 &
 1.98 $\pm$ 0.13 & 1.34 $\pm$ 0.11 & 4.31$\times$0.72 &  18&3 & 8.07$\times$1.10 \\
SEP20-18:06:25 & SEP20-22:13:35 & 3.79 & 2.92 &
 2.40 $\pm$ 0.12 & 1.61 $\pm$ 0.11 & 3.58$\times$0.63 & -20&4 & 6.63$\times$1.14 \\
SEP22-20:08:55 & SEP23-02:16:45 & 5.26 & 3.93 &
 2.22 $\pm$ 0.13 & 1.62 $\pm$ 0.10 & 2.65$\times$0.62 &   9&5 & 4.82$\times$1.14 \\
SEP23-15:07:55 & SEP24-01:38:55 & 9.39 & 6.86 &
 2.41 $\pm$ 0.13 & 1.67 $\pm$ 0.11 & 2.41$\times$0.79 & -45&1  & 4.35$\times$1.44 \\ 
 & & & & 
 2.40 $\pm$ 0.13 & 1.66 $\pm$ 0.10 & 3.22$\times$0.60 &   9&2 & 5.92$\times$1.09 \\
SEP25-20:05:25 & SEP26-00:00:25 & 3.54 & 2.74 &
 2.55 $\pm$ 0.13 & 1.60 $\pm$ 0.11 & 4.11$\times$0.57 &   2&6 & 7.74$\times$1.04 \\
SEP27-16:04:05 & SEP27-19:59:35 & 3.60 & 2.62 &
 2.39 $\pm$ 0.14 & 1.53 $\pm$ 0.11 & 3.30$\times$0.73 & -40&3 & 6.02$\times$1.32 \\
\cline{1-4}
TOTAL          &                &  35.84 & 26.57  \\
\cline{1-4}
& & \multicolumn{2}{r}{average flux} & 2.26 & 1.55 &             & \multicolumn{2}{c}{ } &   \\
& & \multicolumn{2}{r}{median flux}  & 2.39 & 1.60 &             & \multicolumn{2}{c}{ } &   \\
\cline{3-10}
 & & \multicolumn{2}{c}{Complete data set}  & 2.38 $\pm$ 0.09 & 1.64 $\pm$ 0.07 & 1.65$\times$0.73 & 2&6 &  3.01$\times$1.33 \\
\cline{3-10}
\end{tabular}
\end{table*}

\subsection{Reduction}

As the level of variability we expect is small, the reduction needs to
be done as carefully as possible. For this reason, we also 
discuss the reduction in detail.
The data were reduced in Miriad following the user guide (Sault \&
Killeen~\cite{Sault+Killeen99}).
The data were read into Miriad and corrected
for self-interference of the array, as well as
for the phase difference between the X and Y
channels.  Next, the programme {\sc blflag} was
used to flag out bad datapoints interactively.  The calibrators
were then used to determine the antenna gains as a
function of time (using {\sc mfcal}).  

The flux assigned to the flux
calibrator PKS B1934-638 is 5.85 Jy (with a 2~\% error) at 6~cm. The flux
value determined with {\sc uvflux} of the phase
calibrator PKS B0823-500 is 3.0~Jy.
At 3.6 cm the flux calibrator is 2.88 Jy (with a 2~\% error) and
the phase calibrator is 1.53~Jy. 
In those runs where the flux calibrator had not been observed, we
used the phase calibrator with the above mentioned flux values
to calibrate the data.  We determined
the bandpass function from the flux calibrator.

\begin{figure}
\includegraphics{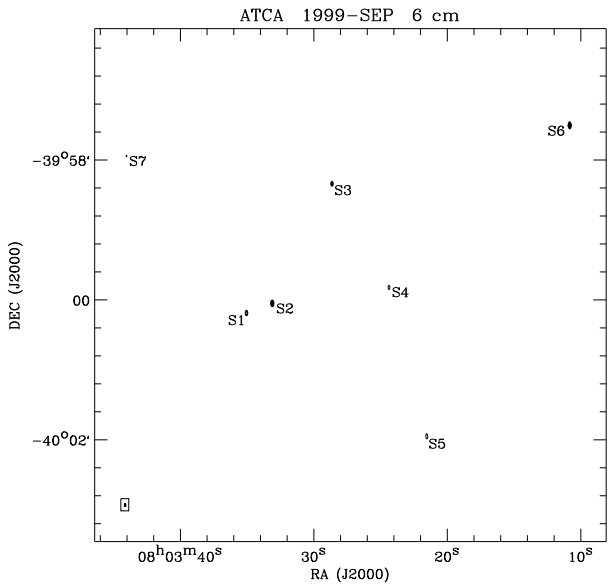}
\caption{$\zeta$ Pup and 6 other sources detected at 6 cm on the ATCA images.
This map covers a large part of the primary beam. 
The synthesized beam is shown in the lower left corner.
Identifications of the sources are listed in 
Table~\ref{table identifications}.
Sources S6 and S7 are
not seen on the 3.6~cm image, because of the smaller primary beam.
}
\label{fig sources}
\end{figure}

We interpolated the instrumental gains and applied them to the $\zeta$~Pup
observation. The task {\sc invert} was then used to produce
an image from the visibility datasets by Fourier transform.
In the inversion we used multi-frequency synthesis (MFS), which
compensates for the spectral index of the source across the bandwidth.
Because of the sparse UV-coverage per observation,
we used robust uniform weighting 
(Briggs~\cite{Briggs95})
to improve the RMS in the map. To de-convolve the image we used
{\sc mfclean}. The resulting clean components are then convolved with
a Gaussian and added to the residual image.
Besides $\zeta$~Pup, 6 other objects were detected in the primary
beam at 6~cm and 4 at 3.6~cm (see Fig.~\ref{fig sources}
and Table~\ref{table identifications}). 

\begin{table}
\caption[]{Position of other sources on the combined ATCA image. 
Formal error bars are better than 0\fm01 in right ascension and
0\farcs1 in declination.
Identifications given refer to Jones~(\cite{Jones85}) and
Condon et al.~(\cite{Condon+al98}). No identification was found for
S4 or S7 in the SIMBAD or HEASARC catalogues. 
}
\label{table identifications}
\begin{tabular}{lllll}
no. & RA (2000) & DEC (J2000) & Identification \\
\hline
S2 & 08 03 33.11 & -40 00 03.1 & \object{EQ 0801$-$398} \\
S3 & 08 03 28.65 & -39 58 20.4 & \object{NVSS J080327$-$395828} \\
\object{S4} & 08 03 24.38 & -39 59 49.2 & \\
S5 & 08 03 21.54 & -40 01 56.9 & \object{NVSS J080321$-$400153} \\
S6 & 08 03 10.87 & -39 57 29.8 & \object{NVSS J080310$-$395729} \\
\object{S7} & 08 03 44.03 & -39 57 56.5 & \\
\hline
\end{tabular}
\end{table}

The cleaning was stopped when the absolute maximum in the residual
map (i.e. the map from which the clean components have been subtracted)
reached 0.30 mJy. This is about 3 to 4 times the theoretical noise of each map,
where the theoretical noise is calculated taking only
the system temperature of the front-end
receiver into account, not the calibration errors, side-lobes or any
other instrumental effects. 
Experience shows that at this cutoff the
RMS in the cleaned map and in the residual map are both
about equal to the theoretical RMS.
At the end, to correct for primary beam
attenuation, the task {\sc linmos} was used.

All images were reduced in exactly the same way with exactly the same
parameters.

\begin{figure}
\includegraphics[scale=0.65]{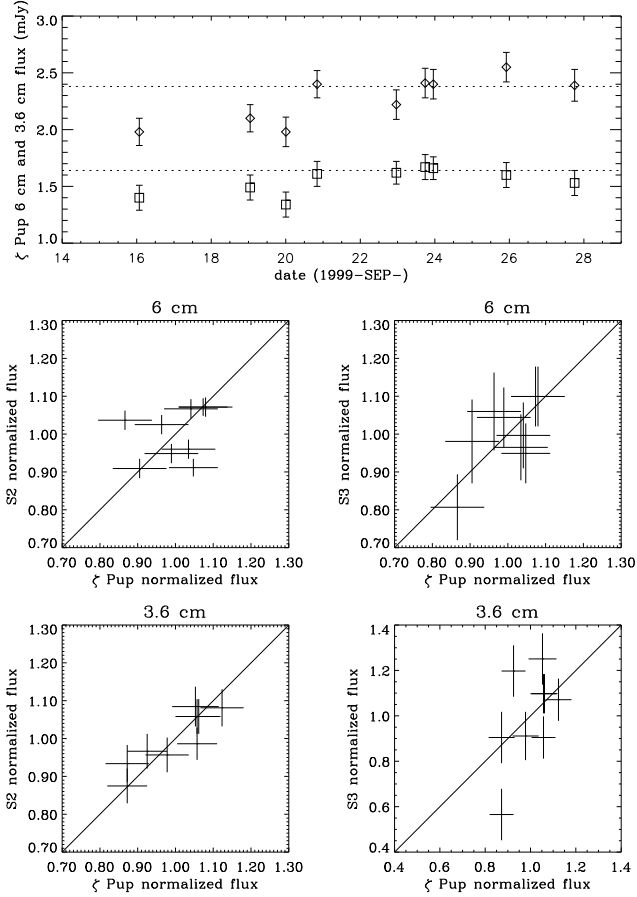}
\caption[]{The top panel shows the $\zeta$~Pup fluxes of each ATCA
observation at 3.6~cm
($\Diamond$) and 6~cm ($\Box$) as a function of time. 
The flux derived from the single dataset combining all the ATCA observations
is given by the dotted line.
The lower 4 panels show the correlation of 
the $\zeta$~Pup fluxes with those of
S2=EQ~0801-398 and S3=NVSS J080327-395828. In the correlation plots
the fluxes were normalized by their average.}
\label{fig ATCA variability}
\end{figure}

\subsection{Fluxes and error bars}

The flux of $\zeta$~Pup was determined by fitting an elliptical
Gaussian, with values for the major and minor axis and position
angle kept fixed to the beam values.
Table \ref{table ATCA} lists the flux values at 3.6 and 6~cm,
and the respective beam sizes with their
position angle. Because the observations at 3.6 and 6~cm were done
simultaneously, the position angles are the same at both wavelengths.
The data obtained on SEP23-24 were split in two sets in order to have the
same time span as in the other observing runs.
The 6~cm fluxes of the other sources in the image are less than 1.2 mJy,
except S2 and S6 which are about 4 mJy (uncorrected for the primary
beam effect): these values are low enough that we do not have to worry
about the effect of their sidelobes.
The fitting procedure ({\sc imfit}) gives an
RMS error on the flux measurement.
To take into account the calibration error, we added 2~\% of the flux values to 
this RMS, and thus arrived at the final error bars
(listed in Table~\ref{table ATCA}).

This error bar only covers the random sources of error.
To get a feeling
for the systematic errors, we redid the reduction in a slightly different
way. Instead of robust uniform weighting, we tried both uniform weighting
(by setting the {\it robust} parameter to $-2$) and natural weighting
({\it robust} parameter = $+2$).
In the latter, the RMS in the
maps is lower at the expense of the more elongated beam and worse
side-lobe levels. In the former, the beam shape is smaller at the
expense of higher RMS levels in the map. 
The robust uniform and natural weighting flux determinations agree 
well within the error bar. As usual, 
the better the UV coverage and beam shape the less difference there is
among the different flux determinations. 
We also measured the flux by determining the maximum intensity of the point
source, instead of fitting an elliptical Gaussian. This alternative flux
determination always falls within the error bar.

\begin{figure}
\includegraphics{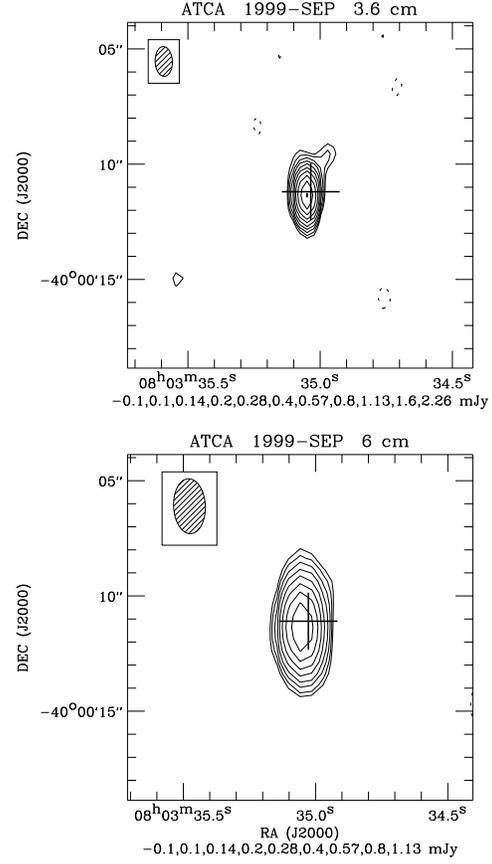}
\caption{ATCA observation (all data combined) of $\zeta$~Pup at 3.6 and 6 cm.
The cross indicates the optical
ICRS 2000.0 position (from SIMBAD), corrected for proper motion
(see Table~\ref{table stellar parameters}).
The contour levels follow a logarithmic scale: their values 
are listed at the bottom of each figure. The negative
contour is given by the dashed line. The first positive contour is at
about three times the RMS noise in the map. The beam is shown in the upper left
corner.}
\label{fig ATCA observations}
\end{figure}

\subsection{Results}

The resulting 3.6 and 6~cm fluxes are plotted in the top panel of
Fig.~\ref{fig ATCA variability}.
While this figure suggests that $\zeta$~Pup is variable at both wavelengths,
we found a similar behaviour in other sources on the map.
We therefore 
compared the flux values for $\zeta$~Pup with two of the brightest
sources close-by (S2=EQ~0801-398 and S3=NVSS J080327-395828, see 
Fig.~\ref{fig sources}). The comparison 
(lower panels on Fig.~\ref{fig ATCA variability})
shows that all three sources follow the same trend. Hence, this argues
strongly against $\zeta$~Pup showing variability.
A further argument against variability follows from the fluxes derived from the
complete data set (see below) which are also plotted on
Fig.~\ref{fig ATCA variability} (dotted lines). Obviously, the flux of
the complete data set is not the straight average of the separate fluxes.
This points to problems in the interpolation of the instrumental gain
phases during the first few runs of the observing session, causing part of the
$\zeta$~Pup flux to be scattered over the rest of the image. These runs
get less weight in the complete data set due to the non-linear nature of
the cleaning process.
Having concluded that the present data do not show detectable variability 
in $\zeta$~Pup, we can derive upper limits on the amount of variability
from the range in fluxes (with their error bars):
these are $\pm 20$~\% of its flux value at 3.6 and at 6~cm.

In principle, the non-detection of variability could be caused by
a bad coverage of the phase for the various periods. However,
the 8 observations
spread over 12 days cover the 5.2 day rotation period 
reasonably well.
We also checked the phase coverage of the variability cycle for 
possible periods 
around 8.5 h and 19 h (see Sect.~\ref{section stellar parameters})
and found it to be good.

As there is no variability, we combined all visibility data 
and Fourier transformed them to obtain a single map.
Because the RMS in this combined map is lower than in each individual map, 
we cleaned it deeper, down to 0.1~mJy.
The resulting images at 3.6 and 6 cm 
are shown in Fig.~\ref{fig ATCA observations}.
The measured fluxes and their error bars (RMS + 2\% of the flux) are listed in
Table~\ref{table ATCA}.

To get a feeling for the systematic errors on the combined data set, 
we investigated the effect of the cutoff limit in the {\sc
clean}ing procedure: we {\sc clean}ed the total map at 6~cm ({\it
robust=0.5}, rms=0.024~mJy) down to 0.07~mJy and 0.15 mJy.  In the
former case the number of clean components was 2769 in the latter 160.
The resulting $\zeta$~Pup fluxes fall well within the error bar.
Also using a single, large, cleaning box instead of multiple cleaning boxes
gave negligible differences in the flux.
Finally, we redid the reduction,
systematically dropping one of the observing runs. The range of values
thus obtained again falls within the error bar.

\section{VLA observations}
\label{section VLA observations}

\subsection{20 cm observation}
\label{section 20 cm observation}

We obtained a 20 cm (L-band) 
observation on 2002 March 27 with the VLA
in A configuration (i.e. the configuration with the highest spatial resolution).
The observation alternated between $\zeta$~Pup and the phase calibrator
\object{0814-356} (J2000). Two runs of 7 min were made on $\zeta$~Pup
and 3 runs of 2 min on the phase calibrator. For the
flux calibration we also observed \object{3C48} = \object{0137+331} (J2000).
The observation consists of two sidebands (at 1.3851 and 1.4649 GHz),
each of 50 MHz bandwidth.

\begin{figure}
\includegraphics{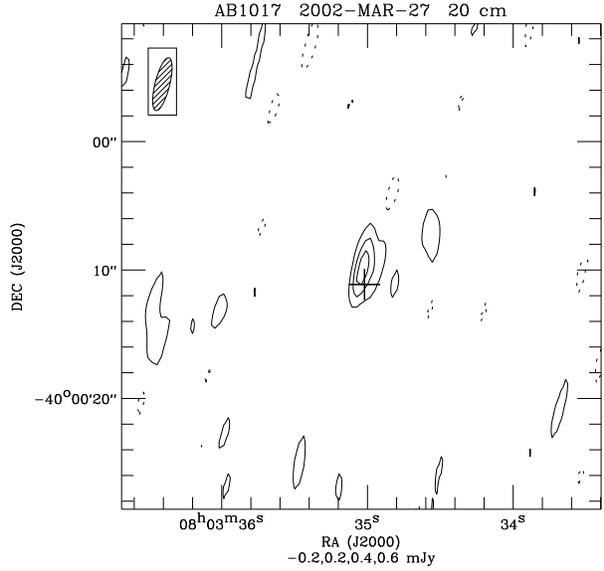}
\caption{VLA observation at 20 cm. 
The contour levels follow a linear scale and the values 
are listed at the bottom of the figure. The negative
contour is given by the dashed line. The first positive contour is at
about twice the RMS noise in the map. The beam is shown in the upper left
corner.
The cross indicates the optical
ICRS 2000.0 position (from SIMBAD), corrected for proper motion.
The $\sim$~1\farcs4 offset with the radio position is most probably
due to ionospheric refraction, which is more important at longer
wavelengths and at the low elevation at which $\zeta$~Pup was observed.
}
\label{fig VLA 20 cm}
\end{figure}

The reduction of these data was done using the NRAO package AIPS 
(Astronomical Image Processing System), following the same steps
as for the ATCA data (Sect.~\ref{section ATCA observations}).
Technical details of the reduction
are listed in Table~\ref{table VLA data reduction}. We stopped cleaning
when the algorithm started finding about the same number of negative as
positive components. The resulting map is shown in
Fig.~\ref{fig VLA 20 cm}.

By fitting an elliptical Gaussian to $\zeta$~Pup,
we found a 20 cm flux of $0.76 \pm 0.09$ mJy. 
The error bar is the RMS noise in the
total map. 
We also added a 2~\% calibration error to the
result (Perley \& Taylor~\cite{Perley+Taylor02}), but this does not
change the resulting error bar significantly.
The maximum intensity of $\zeta$~Pup is 0.74 mJy/beam: the
close agreement with the value from the Gaussian fit
shows that $\zeta$~Pup is indeed
a point source. 

To judge the robustness of our flux determination, we repeated the reduction,
systematically dropping one antenna, using different weightings of distant
visibilities, or natural weighting instead of robust uniform,
or doubling or halving the number of clean components. In all
cases, the results fall within the error bar. The RMS error bar therefore
not only covers the statistical uncertainty, but the systematic errors
as well.

\subsection{Archive observations}
\label{section archive observations}

The VLA archive contains a number of $\zeta$~Pup observations, which
are listed in Table~\ref{table data zeta pup}. 
Many of these observations have not been published previously.
To avoid introducing systematic effects we decided to reduce the whole set.
The procedure is the same as that followed in 
Sect.~\ref{section 20 cm observation}.
The details of the reductions are summarized in 
Table~\ref{table VLA data reduction}. In principle, the 1.3 cm (K-band)
and 2 cm (U-band) observations should be corrected for differential atmospheric 
extinction (using the AIPS task {\sc elint}), 
but in none of those cases was there sufficient information to apply this.

\begin{figure*}
\includegraphics{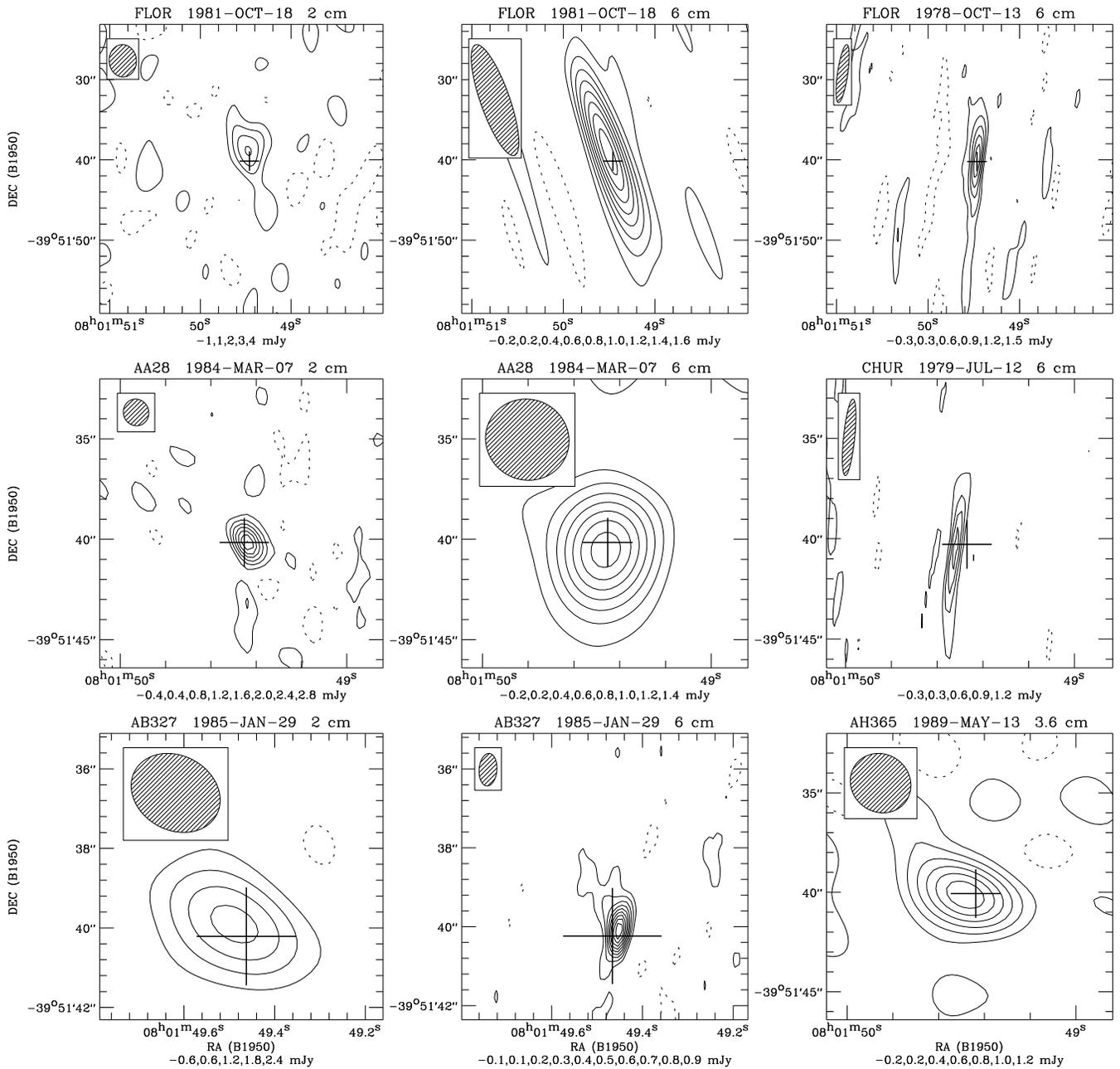}
\caption{Same as Fig.~\ref{fig VLA 20 cm}, but for the VLA archive data
at 2, 3.6 and 6~cm. Only those observations having a detection are shown.
Note the different spatial scales and different contour levels
of the maps.}
\label{fig VLA archive data}
\end{figure*}

The resulting maps are shown in 
Fig.~\ref{fig VLA archive data} and the fluxes are listed
in Table~\ref{table VLA data reduction}. The error bars show the range of results
found by various reductions where we systematically dropped one antenna,
used different weightings of distant
visibilities, or natural weighting instead of robust uniform.
The error bars are always larger than the RMS noise in the map. 
For each observation, we compared the peak intensity of the source to
the flux derived from the Gaussian fit. With the only exception of the 3.6
cm observation of AH365 (see below), there is very
good agreement, showing that the sources are indeed point sources.
When we doubled or halved the number of clean components, 
we were always well within the error bar.
The error bars of the 3.6, 6 and 20
cm observations include a 2~\% calibration error, and the 2 cm a 5~\%
calibration error.

\begin{table*}[t]
\caption{Reduction of the VLA data. Column (1) gives the wavelength,
(2) the programme name, (3) the date of the observation,
(4) the configuration the VLA was in at the time of the
observation, (5) the name of the flux calibrator used, 
(6) the flux (in Jy) assigned to the flux calibrator
(two values are given if there are two sidebands), 
(7) the phase calibrator (J2000 coordinates), 
(8) the number of antennas that gave a usable signal,
(9) the integration time (in minutes)
on the target, (10) the beamsize (major $\times$ minor axis) in arcsec$^2$,
(11) the measured flux (in mJy) and (12) refers to the notes.
The fluxes of the flux calibrators are based on
Perley \& Taylor~(\cite{Perley+Taylor02}).}
\label{table VLA data reduction}
{\small
\begin{center}
\begin{tabular}[]{rllllclrrr@{.}l@{$\times$}r@{.}llc}
\\
\multicolumn{1}{c}{(1)} & \multicolumn{1}{c}{(2)} & \multicolumn{1}{c}{(3)} &
\multicolumn{1}{c}{(4)} & \multicolumn{1}{c}{(5)} & \multicolumn{1}{c}{(6)} &
\multicolumn{1}{c}{(7)} & \multicolumn{1}{c}{(8)} & \multicolumn{1}{c}{(9)} &
\multicolumn{4}{c}{(10)} & \multicolumn{1}{c}{(11)} & \multicolumn{1}{c}{(12)} \\
\multicolumn{1}{c}{$\lambda$} &
\multicolumn{1}{c}{progr.} & \multicolumn{1}{c}{date} &
\multicolumn{1}{c}{conf.} &
\multicolumn{2}{c}{flux calibrator} &
\multicolumn{1}{c}{phase} &
\multicolumn{1}{c}{no.} &
\multicolumn{1}{c}{intgr.} & \multicolumn{4}{c}{beamsize} & 
\multicolumn{1}{c}{flux} & \multicolumn{1}{c}{notes} \\
\multicolumn{1}{c}{(cm)} &
                           &                           &
                           &
\multicolumn{1}{c}{name} & \multicolumn{1}{c}{flux (Jy)} &
\multicolumn{1}{c}{calibrator} &
\multicolumn{1}{c}{ants.} &
\multicolumn{1}{c}{time} & \multicolumn{4}{c}{(arcsec$^2$)} & 
\multicolumn{1}{c}{(mJy)} & \\
\hline
1.3 & FLOR  & 1981-10-18    & C      & \object{3C286} & 2.517       & \object{0828$-$375} & 19 &  56 
     & 5&6  &  3&0             & $<13$           & 1,2 \\
\\
2   & NEWE  & 1979-02-09    &        &  ---  &  ---        & \object{0730$-$116} &  8 &   9 
     & \multicolumn{4}{c}{---} & $<13$           & 1,2,3 \\
    & FLOR  & 1981-10-18    & C      & 3C286 & 3.452       & 0828$-$375 & 24 &  64 
     & 5&1  &  3&5             & 4.3 $\pm$ 0.9   & 1 \\
    & AA28  & 1984-03-07    & CnB    & 3C286 & 3.423/3.432 & 0828$-$375 & 24 &  18 
     & 1&7  &  1&3             & 2.9 $\pm$ 0.3   & 1,4 \\
    & AB327 & 1985-01-29    & A      & \object{3C48}  & 1.742/1.748 & 0828$-$375 & 26 & 121 
     & 3&0  &  1&8             & ---             & 1,5,6 \\
\\
3.6 & AH365 & 1989-05-13    & CnB    & 3C48  & 3.171/3.153 & 0828$-$375 & 26 &  14 
     & 4&0  &  3&0             & 1.4 $\pm$ 0.3   & 6,7 \\
\\
6   & FLOR  & 1978-07-23    &        & 3C286 & 7.462       & \object{0836$-$202} & 11 &   9 
     & \multicolumn{4}{c}{---} & $<4$            & 8 \\
    & FLOR  & 1978-10-13    &        & 3C286 & 7.462       & 0836$-$202 &  9 & 160 
     & 9&4  &  1&3             & 1.7 $\pm$ 0.3   & 9 \\
    & BIEG  & 1978-11-05    &        & 3C286 & 7.462       & 0836$-$202 &  8 &  24 
     &16&   &  0&69            & $<4$            & 8 \\
    & CHUR  & 1979-07-12    &        & 3C286 & 7.462       & 0828$-$375 & 12 & 140 
     & 4&9  &  0&55            & 1.4 $\pm$ 0.3   & 10,11 \\
    & FLOR  & 1981-10-18    & C      & 3C286 & 7.462       & 0828$-$375 & 27 &  28 
     &18&   &  3&6             & 1.71 $\pm$ 0.14 & \\
    & AA28  & 1984-03-07    & CnB    & 3C286 & 7.462/7.510 & 0828$-$375 & 24 &  18 
     & 5&0  &  4&1             & 1.49 $\pm$ 0.11 & 12 \\
    & AB327 & 1985-01-29    & A      & 3C48  & 5.405/5.459 & 0828$-$375 & 27 &  91 
     & 1&1  &  0&45            & 1.05 $\pm$ 0.08 & 6 \\
\\
20  & FLOR  & 1979-02-16    &        & 3C286 & 14.51       & 0836$-$202 & 11 & 182 
     &14&   &  2&3             & $<1.5$          & 2 \\
    & FLOR  & 1981-10-18    & C      & 3C286 & 14.51       & 0828$-$375 & 26 &  18 
     &60&   & 12&              & $<0.75$         & 2 \\
    & AB1017& 2002-03-27    & A      & 3C48  & 15.49/16.20 & \object{0814$-$356} & 26 &  14 
     & 5&5  &  1&1             & 0.76 $\pm$ 0.09 & \\
\hline
\end{tabular}
\end{center}

Notes:\\
\begin{tabular}{rl}
1 & Insufficient data for atmospheric extinction correction \\
2 & Upper limit from 3 sigma \\
3 & No flux calibrator, used 0532+075 (J2000) 
    instead with an assumed flux of 2.20 Jy \\
4 & Bieging et al.~(\cite{Bieging+al89}) list 3.0$\pm$0.2 mJy 
     for this observation \\
5 & Use of 3C48 as a flux calibrator in the A configuration is not recommended 
    (Perley \& Taylor~\cite{Perley+Taylor02}) \\
6 & Data reduction problems detailed in Sect.~\ref{section archive observations} \\
7 & Lamers \& Leitherer~(\cite{Lamers+Leitherer93}) list 1.60$\pm$0.07 mJy 
    for this observation, based on work by Howarth \& 
    Brown~(\cite{Howarth+Brown91}) \\
8 & Upper limit derived from non-detection of EQ 0801-398 \\
9 & Primary is resolved; true flux might be slightly lower \\
10 & An additional 10~\% uncertainty in the flux calibration should be added,
     because the instrumental gains of the flux and phase calibrator \\
   & show significant differences (Fomalont \& Perley~\cite{Fomalont+Perley99}) \\
11 & Abbott et al.~(\cite{Abbott+al80}) and Bieging et al.~(\cite{Bieging+al89}) 
    list 1.4$\pm$0.3 mJy for this observation \\
12 & Bieging et al.~(\cite{Bieging+al89}) list 1.3$\pm$0.1 mJy 
     for this observation \\
\end{tabular}
}
\end{table*}

As $\zeta$~Pup is rather low on the horizon from the VLA, the quality of the
observations is not always good. 
The AB327 observations at 2 and 6 cm show
large variations in the gain phases
(30-50 degrees) of the calibrators. All sources on the 6 cm image
(including $\zeta$~Pup) are 30-50~\% lower in flux compared to other observations.
The 2 cm $\zeta$~Pup flux is heavily
dependent on tapering, and is
compatible with a value of $\sim 3$~mJy.
The 3.6 cm observation of AH365 is the only one where
the measured peak intensity is systematically higher (by $\sim 40$~\%) than the
flux derived from the Gaussian fit. This suggests that,
due to phase problems, the source is not quite a point source.
The measured peak intensity is 2 mJy.
Minor comments on other observations 
are given in Table~\ref{table VLA data reduction}.

In general, the agreement of the fluxes we derived
with published values is quite good (see notes to 
Table~\ref{table VLA data reduction}). Only for the AA28 - 6 cm observation
do we find a value (1.49 $\pm$ 0.1 mJy) significantly higher than
the published value (1.3 $\pm$ 0.1 mJy, Bieging et al.~\cite{Bieging+al89}).

There is an additional
observation available (AC308 -- see Table~\ref{table data zeta pup})
made at 20 cm. This observation is part of the NRAO VLA Sky Survey
(NVSS -- Condon et al.~\cite{Condon+al98}). We checked this survey and
found that $\zeta$~Pup was not detected (but EQ~0801-398 is detected).

\subsection{Long term variability}
\label{section long term variability}

\begin{figure*}
\includegraphics[scale=1.0]{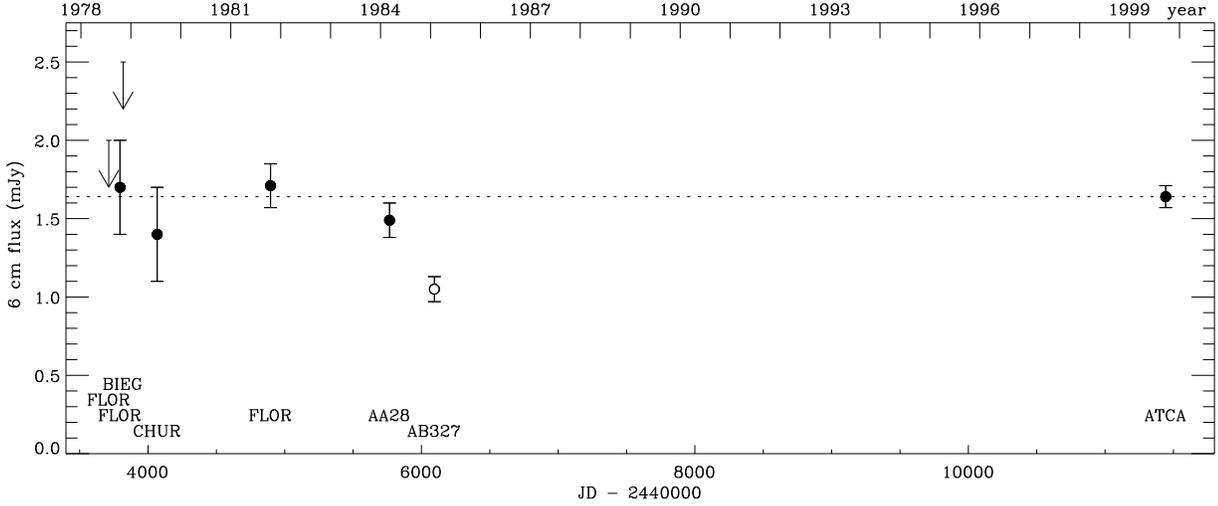}
\caption[]{The 6 cm fluxes of $\zeta$~Pup as a function of time.
The open circle indicates an observation (AB327) that might be unreliable (see
Sect.~\ref{section archive observations}). The dotted line shows
the flux from the ATCA combined data.
}
\label{fig 6 cm}
\end{figure*}

We now compare the different VLA observations, wavelength by wavelength,
to see if we can detect variability at timescales much longer than
the rotation period.

We have only two 2 cm flux determinations. The error bars do not overlap,
which might suggest variability. However, we recall the considerable
difficulties in the reduction of these data
(Sect.~\ref{section archive observations}), and based on that, we do
not consider them to present evidence of variability.
For completeness, we also mention the 2 cm observations of $\zeta$ Pup by
Morton \& Wright (\cite{Morton+Wright79}): they list 7.2$\pm$1.1 mJy
as the average of two observing runs. This value
is significantly higher than ours. However, these observations
were made with a single-dish antenna (Parkes 64-m), which has a beam
of 2\farcm3. When targeting $\zeta$~Pup, the beam
also covers S2=EQ 0801-398 (see Fig.~\ref{fig sources}), which
will dominate the measured flux. These data can therefore not be used
to look for variability of $\zeta$~Pup.

The single VLA 3.6 cm observation ($1.4 \pm 0.3$ mJy)
can be compared to the ATCA determination ($2.38 \pm 0.09$ mJy).
In Sect.~\ref{section archive observations} we showed that the AH365 observation
was quite problematic.
Comparing the ATCA measurements of sources 
S2=EQ 0801-398 and S3=NVSS J080327-395828 to the AH365
measurements shows only a 20~\% effect for S2, and none for S3, suggesting
that the Gaussian fit fluxes would be reliable. In that case, the
AH365 $\zeta$~Pup 
flux would be significantly different from the ATCA determination.
However AH365 shows such an accumulation of problems (the flux of the
secondary calibrator is
considerably less than expected, there are reasonably high phase changes
and there is the 
difference between peak intensity and Gaussian integrated flux) that 
we do not consider this convincing evidence of variability.

The single 20 cm determination is compatible with the two upper limits.

The only wavelength for which we have a reasonable number of flux determinations,
is 6 cm. Fig.~\ref{fig 6 cm} shows these fluxes as a function of time.
The flux from the ATCA combined data
goes through the error bars of almost all observations,
indicating that there is no detectable variability. 

The only exception
is the significantly lower flux of AB327. 
In Sect.~\ref{section archive observations} we discussed the phase problems
of this observation, which lead to the lower flux. 
Another possible factor might be that we are starting to resolve the stellar
wind of $\zeta$~Pup (this is the observation with the highest spatial
resolution we have).
To estimate how much flux we would lose, we use a Wright \& Barlow 
(\cite{Wright+Barlow75}) approach. By their definition of the characteristic
radius ($R_\nu$) of the radio-emitting region, the observed flux 
($S_\nu$) is given by:
\begin{equation}
S_\nu = 10^{-26} \int_{R_\nu}^{+\infty} {\rm d}r \frac{4 \pi r^2}{D^2} K(\nu,T) \gamma 
   \left( \frac{\dot{M}}{4 \pi r^2 v_\infty \mu m_{\rm H}} \right)^2 B_\nu (T),
\end{equation}
where $r$ is the radius, $D$ the distance to the star, $K(\nu,T)$ the
free-free absorption coefficient, $\gamma$ the ratio of electron to ion number
densities, $\dot{M}$ the mass loss rate, $v_\infty$ the terminal velocity,
$\mu$ the average atomic mass, $m_{\rm H}$ the proton mass and $B_\nu (T)$
the Planck function at frequency $\nu$ and temperature $T$. All units are
cgs, except for $S_\nu$ which is in mJy.
If, instead of integrating to $+\infty$, we only integrate to a radius
$R_{\rm max}$, we lose a fraction of the flux given by:
\begin{equation}
\frac{\Delta S_\nu}{S_\nu} = \frac{R_\nu}{R_{\rm max}}.
\end{equation}
The beamsize ($\theta_{\rm beam}$) determines $R_{\rm max}$ by:
$\theta_{\rm beam} = 2 R_{\rm max}/D$. Combining this with the Wright \& Barlow
expression for $R_\nu$, we get:
\begin{equation}
\frac{\Delta S_\nu}{S_\nu} = 3.8 \times 10^5
   \frac{(\gamma g Z^2)^{1/3} T^{-1/2}}{D \theta_{\rm beam}}
  \left( \frac{\dot{M} \lambda}{\mu v_\infty} \right)^{2/3},
\end{equation}
where $\dot{M}$ is in $M_{\sun}/\mathrm{yr}$, $v_\infty$ in km/s, $\lambda$ in cm,
$\theta_{\rm beam}$ in arcsec, $D$ in kpc. We
take $\dot{M} = 5.9\times 10^{-6}$ (Puls et al.~\cite{Puls+al96}),
$\mu$=1.4, $\gamma$=1, $Z$=1 and the Gaunt-factor $g$=7. Other parameters
are taken from Table~\ref{table stellar parameters}. With a beam of
$\theta_{\rm beam}$=0{\farcs}7, we find a flux loss of $6-12$~\%,
depending on whether we take a hot wind (42\,000 K) or a cool wind (10\,000 K).
If we use the mass loss rate derived from our model for the radio observations
(Table~\ref{table mass loss rate determinations}), we find a $4-8$~\% effect.

The low AB327 flux can thus not be explained by
our starting to resolve the stellar wind, but is due
to problems with the interpolation of the gain phases.

\section{JCMT observation}
\label{section JCMT observations}

We also determined the 850 $\mu$m flux of $\zeta$~Pup using the
Submillimetre Common-User Bolometer
Array (SCUBA, Holland et al.~\cite{Holland+al99}) on JCMT
(see Table~\ref{table data zeta pup}).
The instrument observes
simultaneously at 450 and 850 $\mu$m using two hexagonal arrays of bolometers.
The sensitivity at 450 $\mu$m is too low for a detection, so we will only
discuss the 850 $\mu$m data.
Our flexibly scheduled observations were taken
in medium weather conditions (850 $\mu$m zenith opacity was 0.32 -- 0.47)
during a half-shift on 2000 October 12. The beam at 850 $\mu$m
is 14\farcs5.

As the data were collected in the same run as the $\epsilon$~Ori 
observation discussed 
by Blomme et al.~(\cite{Blomme+al02}), we refer to that paper for
details on the observation and the reduction.
The total on-target integration time for $\zeta$~Pup was 15 min.
After reduction, the RMS on the $\zeta$~Pup
observation turns out to be about 16~\%.
We tried variants in the reduction
(e.g. using other bolometers than the inner ring for sky-noise removal),
but this changes the flux by considerably less than 16~\%.

The $\zeta$~Pup observations are preceded and succeeded by an observation
of the calibrator \object{OH231.8}. Taking the average of these two calibration
observations, we arrive at $31 \pm 5$~mJy for the flux.
The error bar takes into account the measurement errors on the target
and the calibrator as well as the calibration error. However,
OH231.8 is somewhat variable and the flux of the calibrator at the time of our
observation could be different from the reference flux we
used.

If we use the \object{HL Tau} calibration observation that precedes our run,
we have $28 \pm 5$~mJy.
Trying to use other
calibrators that were observed during that night (but of course further away
in time from our observation) tends to favour the lower value.
In view of this we propose $28 \pm 5$~mJy as the best determination
of the flux. The $\zeta$~Pup value is then also based on the same
calibrator as the $\epsilon$~Ori observation analysed by
Blomme et al.~(\cite{Blomme+al02}).

\begin{table}
\caption{Submillimetre and radio fluxes of $\zeta$~Pup. Data from this paper,
unless otherwise indicated.}
\label{table all fluxes}
\begin{tabular}{lllll}
$\lambda$ & flux (mJy)     & reference \\
\hline
850 $\mu$m & 28   $\pm$ 5    & JCMT \\
1.3 mm     & 20.2 $\pm$ 1.8  & from Leitherer \& Robert~(\cite{Leitherer+Robert91})\\
2 cm       & 4.3  $\pm$ 0.9  & VLA, FLOR 1981-10-18 \\
2 cm       & 2.9  $\pm$ 0.3  & Bieging et al.~(\cite{Bieging+al89}), revised \\
3.6 cm     & 2.38 $\pm$ 0.09 & ATCA \\
6 cm       & 1.64 $\pm$ 0.07 & ATCA \\
20 cm      & 0.76 $\pm$ 0.09 & VLA \\
\hline
\end{tabular}
\end{table}

If we want to compare our 850 $\mu$m value with the 1.3 mm determination
of Leitherer \& Robert~(\cite{Leitherer+Robert91}), we need to correct
for the difference in wavelength. If we assume an $\alpha=0.6$ spectrum,
we find that our value corresponds to $22 \pm 4$ mJy at 1.3 mm, which agrees
very well with the Leitherer \& Robert value of $20.2 \pm 1.8$~mJy.
Our value is also compatible with the
Altenhoff et al.~(\cite{Altenhoff+al94}) upper limit of 33 mJy at 1.2 mm.

\section{Discussion}
\label{section discussion}

\subsection{Smooth wind model}
\label{section smooth wind model}

Table~\ref{table all fluxes} summarizes the submillimetre and radio
fluxes we will discuss in this section.
Other fluxes either agree with these, or possible disagreements
have been explained in Sect.~\ref{section archive observations}.

From the 1.3 mm and 6 cm fluxes we can derive the spectral
index $\alpha$ (defined by $F_\nu \propto \lambda^{-\alpha}$).
The measured value of $\alpha = 0.66 \pm 0.03$ defines a power law that
goes through the error bars of all observations, with the exception of
the 2 cm fluxes (where it is intermediate between the two determinations). 
This spectrum is
somewhat steeper than expected from the Wright \&
Barlow (\cite{Wright+Barlow75}) model (where $\alpha = 0.6$),
indicating some discrepancy between the millimetre and radio fluxes.

To better quantify this discrepancy, we also
made a smooth wind model for $\zeta$~Pup in the same
way as in Runacres \& Blomme (\cite{Runacres+Blomme96}).
The model solves the equations of radiative transfer and statistical
equilibrium in a spherically symmetric stellar wind, containing
only hydrogen and helium. The density in the
wind is determined by solving the time-independent hydrodynamical
equations (following Pauldrach et al.~\cite{Pauldrach+al86}). When
fitting the model to the observations, the visual and near-infrared
fluxes were used to determine the interstellar extinction and the radio
fluxes to determine the mass loss rate. The far-infrared and millimetre
fluxes are unconstrained and can therefore be used to see how well
the smooth wind model fits the observations. Further details of the
model are given in Runacres \& Blomme. A new version of their 
$\zeta$~Pup model was calculated
that uses the parameters listed in Table~\ref{table stellar parameters}.

In fitting the model, we fixed the mass loss rate from our ATCA 6 cm
observation. We choose this observation because it has the smallest
error bar. The best fit gives a mass loss rate of $\dot{M}$ = 
$3.5 \pm 0.1 \times 10^{-6}$ $M_{\sun}$/yr.
The error bar on
$\dot{M}$ corresponds only to the error bar on the 6~cm flux,
and does not include the more important 
errors due to stellar parameters or distance.
E.g., the error in the Hipparcos distance converts to a ($-0.9$, $+1.5$)
$\times 10^{-6}$ $M_{\sun}$/yr
error on $\dot{M}$.

\begin{figure}
\includegraphics[scale=0.55]{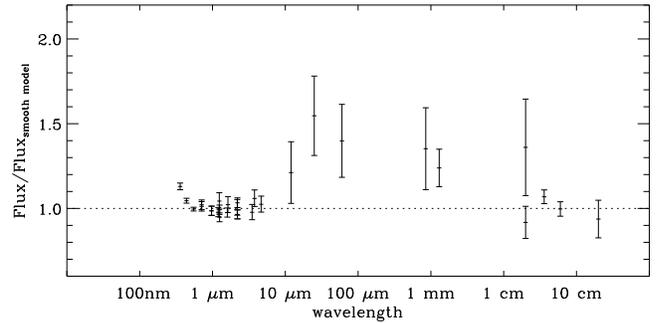}
\caption{Observed fluxes normalised to a smooth wind model. Observations
above the dotted line point to additional emission that is not included
in the smooth model.}
\label{fig smooth model}
\end{figure}

\begin{table}[t]
\caption{Comparison to other mass loss rate determinations of
$\zeta$~Pup.}
\label{table mass loss rate determinations}
{\small
\begin{tabular}{llll}
\multicolumn{1}{c}{wavelength} & \multicolumn{1}{c}{$\dot{M}(10^{-6} M_{\rm \sun}/{\rm yr})$} &
   \multicolumn{1}{c}{reference} \\
\hline
radio    & $2.4_{-0.7}^{+1.0}$ & Lamers \& Leitherer~(\cite{Lamers+Leitherer93}) \\
H$\alpha$& $3.5_{-1.2}^{+1.9}$ & Lamers \& Leitherer~(\cite{Lamers+Leitherer93}) \\
         & 5.9 & Puls et al.~(\cite{Puls+al96}) \\
radio    & 3.5 & this paper \\
\hline
\end{tabular}
}
\end{table}

\begin{figure*}
\includegraphics[scale=1.0]{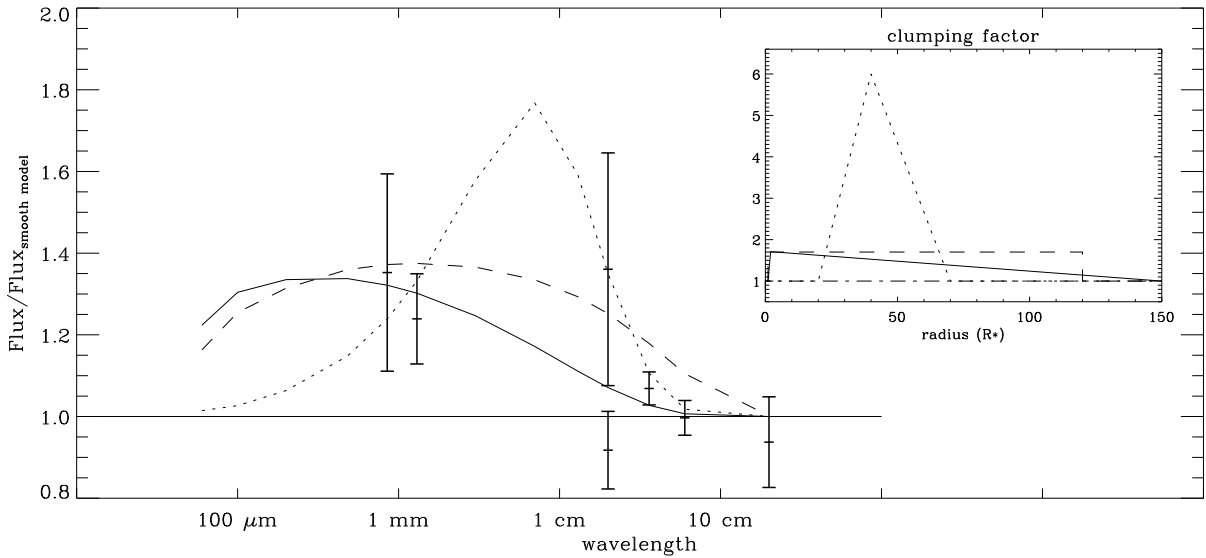}
\caption{The millimetre excess is fitted with a structured wind model. Three
possible runs of the clumping factor as a function of radius are used
(inset). The dashed line presents a model that has too much structure too
far out in the wind to be able to explain the observations.}
\label{fig clumped model}
\end{figure*}

Because it is such a well-studied star, a large number of mass loss rate
determinations have been made for $\zeta$~Pup. In our comparisons we
will only consider the more recent determinations
(Table~\ref{table mass loss rate determinations}). Our radio mass loss rate
of $3.5 \times 10^{-6}$ $M_{\sun}$/yr
is different from the $5.9 \times 10^{-6}$ $M_{\sun}$/yr
value by Puls et al.~(\cite{Puls+al96}) based on fitting the
H$\alpha$ line profile.
This difference is significant, considering that we used the same stellar 
parameters as they did. 
We recall that Petrenz \& Puls~(\cite{Petrenz+Puls96}) already noted
this discrepancy and tried to explain it
by invoking an equatorial density enhancement, caused by
the large rotational velocity of the star.
Our results are in acceptable agreement with both the radio and
H$\alpha$ determinations by Lamers \& Leitherer~(\cite{Lamers+Leitherer93}),
but we note that they used a simpler model for the H$\alpha$ emission
than Puls et al.
Contrary to what the Lamers \& Leitherer numbers suggest,
$\zeta$~Pup is therefore an exception to the usually
good agreement between the H$\alpha$ and radio mass loss rates.

The observed fluxes are compared to the best-fit smooth wind model in 
Fig.~\ref{fig smooth model}. The figure shows the millimetre and
radio observations listed in Table~\ref{table mass loss rate determinations},
supplemented by visual and infrared data listed in
Runacres \& Blomme~(\cite{Runacres+Blomme96}).
As found by Runacres \& Blomme, 
there is an excess flux at millimetre
wavelengths. Due to our somewhat higher 6~cm flux, the effect
found here (24~\%) is somewhat lower than what they found (35~\%).
The present result is significant at the 2 sigma level.
In view of the large error bars, it is not clear what happens at 2~cm,
but the 3.6~cm flux could be marginally in excess. The 20~cm flux is in
good agreement with the smooth wind model. We stress that the same
discrepancy is found when a much simpler Wright \& Barlow
(\cite{Wright+Barlow75}) model is used instead of our smooth wind model.

\subsection{Explaining the millimetre excess}

The millimetre excess is the most important result from this work.
Various explanations will be considered in this section. To study the excess
we will make considerable use of models similar to
the one developed by Wright \& Barlow (\cite{Wright+Barlow75}).

We start by noting that the Wright \& Barlow formula for the
flux is not very sensitive to temperature, so a radial gradient
of the temperature will not explain the excess. There could
be an indirect effect of the temperature however, due to the recombination
of important ions. This explanation has been proposed by Leitherer \& Robert
(\cite{Leitherer+Robert91}). As He$^{++}$ will recombine much
more quickly than H$^{+}$, we only consider the recombination of 
He$^{++}$.
A careful evaluation of the
ionization-dependent factors in the Wright \& Barlow (\cite{Wright+Barlow75})
formula for the radio flux shows that, if we assume all helium to be
He$^{++}$  in the millimetre formation region and all helium to be
He$^{+}$ in the centimetre formation region,
the discrepancy between the fluxes can be explained.
It should be noted that our own 
smooth-wind
model does not show recombination,
but it does not include important effects, such as line blanketing
due to metals.
Recent models of early-type stars arrive at lower effective temperatures
than the one we used here (Bianchi \& Garcia~\cite{Bianchi+Garcia02},
Bianchi et al.~\cite{Bianchi+al03}, Martins et al.~\cite{Martins+al02}),
which would favour recombination.
Further evidence that recombination occurs, comes from models of the stellar 
wind of $\zeta$~Pup (Hillier et al.~\cite{Hillier+al93}, 
Pauldrach et al.~\cite{Pauldrach+al01}, Puls 2003, pers. comm.).
Due to the recombination in these models, it is necessary
to assume that the X-rays are formed out to large distances in the wind
($> 100~R_*$), because X-rays formed too close to the star get absorbed.
However, the X-ray spectral lines of $\zeta$~Pup 
(Cassinelli et al.~\cite{Cassinelli+al01},
Kahn et al.~\cite{Kahn+al01}) seem to require formation closer
to the stellar surface and less opacity in the wind 
(Kramer et al.~\cite{Kramer+al03}). This fact is difficult to reconcile
with He recombination.

A slower velocity law has been claimed to explain the {\it far-infrared}
discrepancies seen on Fig.~\ref{fig smooth model} 
(Kudritzki \& Puls~\cite{Kudritzki+Puls00}). To explain the {\it millimetre}
discrepancies would require a very much slower velocity law. To investigate
this we adapted the Wright \& Barlow (\cite{Wright+Barlow75}) model to
include a $\beta$-type velocity law. In such case, an analytical solution
is no longer possible, and all integrations were done numerically.
We find that $\beta \approx 5$ is needed to explain the millimetre
fluxes, which is not compatible with other observational indicators such 
as the H$\alpha$ profile, which requires $\beta \approx 1.15$
(Puls et al.~\cite{Puls+al96}).

The existence of a flux excess at millimetre wavelengths can also
be due to structure in the wind.
As there is a wealth of evidence for structure in these winds,
it seems obvious to explore a model based on that.

The model in question is the same as used in Blomme et al.~(\cite{Blomme+al02})
for $\epsilon$~Ori.
The smooth wind opacity in a Wright \& Barlow~(\cite{Wright+Barlow75}) 
model is multiplied by the clumping factor to get the clumped wind opacity. 
The clumping factor is defined as
\mbox{$<\rho^2>$}/\mbox{$<\rho>^2$}, where the symbol
\mbox{$<>$} stands for a time-average, which we have approximated by integrating
over a small volume of the wind.
Optical depth, emergent intensity and flux are then calculated
as in Wright \& Barlow. As our model allows the
clumping factor and velocity to change as a function of distance, all
integrations have to be done numerically.
We used a run of the clumping factor based on the work of
Runacres \& Owocki~(\cite{Runacres+Owocki01}), who calculated time-dependent
hydrodynamical models to study the effect of the line-driving instability at
large distances from the star. The clumping
factor in their models rises to reach a maximum rather far away from the star
($10 - 50~R_*$) and then decreases again.
If we simplify their results somewhat, we can approximate
the typical behaviour for the clumping factor
by a piece-wise linear curve fixed by
specifying three points. We let the clumping
factor rise linearly from close to the surface of the star (point 1)
to a certain distance (point 2), and
then let it fall off again linearly till it becomes one
(point 3). The inset to Fig.~\ref{fig clumped model} shows an example
(dotted line).
We also used piece-wise linear curves fixed by four points.
By setting the clumping factor equal to one at large distances, we ensure
that there is no excess at radio wavelengths.

Fig.~\ref{fig clumped model} shows a number of fits we attempted. Although
no unique solution is found, it is quite clear that the clumping factor has
to be significantly higher than 1 in the inner part of the wind,
and needs to diminish considerably beyond 70~$R_*$. 
To stress this last point, Fig.~\ref{fig clumped model} includes a model
with substantial clumping up to 120~$R_*$ (dashed line); the fluxes clearly
overshoot the 3.6 and 6 cm observations.
The fact that structure diminishes beyond a certain distance
is similar to what we found for
$\epsilon$~Ori (Blomme et al.~\cite{Blomme+al02}), 
except that for $\epsilon$~Ori, this started happening around
$\sim 40~R_*$.

While structure diminishes around 70~$R_*$, it  does not necessarily
disappear. For $\epsilon$~Ori we showed that a model where
structure persists up to large distances (with a constant clumping factor)
could equally well explain the observations, provided we reduce the
mass loss rate accordingly. For $\zeta$~Pup, we see that
the 20 cm observation is in reasonable agreement with the 6 cm one. 
This shows that,
{\em if} there is still clumping left, it falls off considerably slower
than it does in the inner 70~$R_*$ of the wind. Obviously, a more accurate
20 cm flux and observations in the 1 mm -- 3.6 cm region will further
constrain the extent and amount of structure.

The above model was very much inspired by the small-scale structure expected
from the instability of the radiative driving mechanism. However, CIRs,
a disk or a polar enhancement should have similar effects on the millimetre
flux. In these cases the 70~$R_*$ radius will be significant as well,
in that
it shows where the structure diminishes substantially, or maybe even disappears.
A detailed comparison of models for different types of structure is
beyond the scope of the present paper.

\section{Conclusions}
\label{section conclusions}

Radio observations of $\zeta$~Pup covering about two rotational periods,
supplemented by archive observations covering a much longer time scale,
do not show variability at more than the $\pm 20$~\% level. The long integration
time gives us an accurate flux determination of 
2.38 $\pm$ 0.09 mJy at 3.6 cm and 1.64 $\pm$ 0.07 mJy at 6 cm.
These values are slightly higher than the ones previously known.

Converting the fluxes into a mass loss rate, we find $\dot{M}$ = 
$3.5 \times 10^{-6}$ $M_{\sun}$/yr.
This value confirms the significant discrepancy with the H$\alpha$ mass loss
rate (Petrenz \& Puls~\cite{Petrenz+Puls96}).

A smooth wind model shows that the millimetre fluxes are too high compared to the
radio fluxes. While recombination of helium in the outer wind cannot be discounted
as an explanation, we favour a model that ascribes the discrepancy to structure.
A simple model shows a substantial decay, or maybe
disappearance, of structure beyond 70~$R_*$. Fig.~\ref{fig clumped model}
shows how observations at wavelengths between 1 mm and 3.6 cm can further
constrain models for structure.

The present data do not allow a distinction between the various types
of structure (stochastic, CIRs, disk or polar enhancement). Attempting to
detect variability at far-infrared or millimetre wavelengths should provide
better constraints on the azimuthal symmetry of the structure, thereby
allowing us to decide whether it is in the form of CIRs.

\begin{acknowledgements}
We thank Joan Vandekerckhove for his help with the reduction of the
VLA data.
We are grateful to Thomas Lowe for making the JCMT observation.
We also thank the original observers of the VLA archive data we used.
We thank Joachim Puls for information about the helium recombination
in the stellar wind models.
This work benefitted from discussions with Stan Owocki.
This research has made use of the SIMBAD database, operated at CDS, 
Strasbourg, France and NASA's Astrophysics Data System Abstract Service.
We also consulted the High Energy Astrophysics 
Science Archive Research Center (HEASARC), provided by NASA's Goddard Space 
Flight Center.
M.C.R. acknowledges support from ESA-Prodex project no. 13346/98/NL/VJ(ic),
financed by ESA-Prodex.
Part of this research was carried out in the framework of the
project IUAP P5/36 financed by the Belgian State, Federal Office for
Scientific, Technical and Cultural Affairs.
\end{acknowledgements}

\end{document}